# Out-of-Step Detection Based On an Improved Line Potential Energy Criterion

Xiao Li, Chongru Liu, Jin Ma,

*Abstract*—The line potential energy in the cutset is used as the criterion for monitoring the generator instability, but the criterion has the following two limitations due to narrowly defined conditions. The assumption of an ideal constant power load is difficult to satisfy, and the condition of the critical cutset is too narrow. The limitations are addressed by analyzing the relationship between the power of the load elements and the power of the cutset based on the two-machine equivalent. Modifications of the theoretical derivation steps of this criterion are presented to extend the applications of the criterion and provide theoretical support for its online use. Two simple strategies are proposed to ensure early detection of the generator instability.

**Nomenclature**
**Constants**
$S$      the set of leading generators
$A$      the set of lagging generators
$M_i$      the inertia of the $i$-th generator
$P_i^s$      the $P_i$ at a stable equilibrium point after failure.

**Variables**
$\omega_i$      the speed deviation of the $i$-th generator from the nominal value in the center of inertia reference
$\omega_2$      the speed difference between two equivalent clusters
$\delta_i$      the rotor angle of the $i$-th generator in the center of inertia reference
$\sigma_i$      the voltage phase angle difference of the $i$-th electrical components of the lines or with respect to the ground.
$P_i$      the active power of the $i$-th generator, branch, or load

**Energy components**
$V_{KE}$      the total kinetic energy of the generators in the system
$V_{PE}$      total LPE in the system
$V_{KE2}$      relative kinetic energy between two groups
$V_{PE2}$      relative LPE between two groups
$\Delta P_{L2}$      difference between the equivalent load power and the steady-state value of the groups
$\Delta P_C$      the difference in the sum of the cutset lines' power from the leading group to the lagging group and the steady-state value

Xiao Li, and Chongru Liuare with the State Key Laboratory for Alternate Electrical Power System with Renewable Energy Sources, North China Electric Power University, Beijing 102206, China (e-mail: lixiao000@ncepu.edu.cn; chongru.liu@ncepu.edu.cn,).
Jin Ma is with The University of Sydney.

## I. INTRODUCTION

The stable operation of the power system is an important aspect of social development. Due to economic reasons, the operation of the power system may be close to the stability limit. Since the power system inevitably experiences various disturbances or faults, transient instability of the generators or blackouts may occur in severe cases. Therefore, the prevention of the generators' transient instability, also called the out-of-step condition, is very important for a power system.

This problem has attracted people's attention early on. Research on power system stability has a long history, and significant progress has been made in online applications. Traditional methods fall into two categories, i.e., time-domain simulations and energy function methods. The time-domain simulation method has been applied online due to the development of rapid simulation technology [1]. The advantage of the energy function method is its theoretical strength, but it is difficult to apply to complex control models. Many types of energy function methods have been developed to meet the needs of online monitoring [2–5]. Reference [6] developed a measurement-simulation hybrid method suitable for online stability analysis that combined the time domain-simulation and energy function methods. Due to the rapid development of artificial intelligence in recent years, online monitoring methods of transient stability based on artificial intelligence have been increasingly used in recent years [7–9]. However, due to the high degree of dependence on training data, the algorithm stability is uncertain for online power system applications. References [10,11] used the concepts in the energy function method to improve the limitations of artificial intelligence-based methods. Phasor measurement unit (PMU)-based model-free methods are also suitable for online applications due to the relatively low computational complexity [12–14]. However, this method may need to be combined with other methods to ensure reliability [15].

Despite many achievements in the online detection of generator out-of-step conditions, several challenges remain: 1) most methods based on generator measurements rely on the motion information of all generators. Some of these methods are limited to the swing of a single machine with respect to the other groups. 2) It is difficult to ascertain the physical nature of the influence of the network on transient stability because of the subordinate status of the network structure variables and the operational variables in the mentioned methods.

Several rotor angle instability indicators based on cutset measurements and not requiring motion information were proposed to address these challenges. The cutset stability criterion [16] was originally derived from the topological energy-function [17], and its accuracy was improved in a subsequent study [18]. Because this indicator depends on local dynamic measurements, it is easy to use for online detection [19,20] and represents an improvement over detection indicators previously applied to simple systems [21]. However, the theoretical derivation conditions of this type of line potential energy (LPE) criterion are often too narrow. The assumption that the energy level is higher in the cutset lines is not satisfied because of the existence of the voltage-dependent load. In addition, the theoretical derivation focuses on the critical cutset rather than the common cutset, reducing the flexibility of the online use of these indicators.

This paper aims to analyze the applicable conditions of the LPE for out-of-step detection and improve the above limitations. This paper provides the following two contributions:

a) The contribution of the load to the LPE is analyzed. The results demonstrate that the reliability of the LEP criterion comes at the expense of delayed detection of the generator instability. A simple method based on monitoring parts of the load power is developed to compensate for the weakness of the LPE criterion regarding the delayed detection of the instability.

b) The constraint for choosing the cutset lines in the LPE instability criterion is relaxed. The analysis shows that the LPE criterion is not restricted to critical cutsets. A simple strategy is to choose the cutset lines close to the low inertia group to improve instability monitoring.

The remainder of this paper is organized as follows. Section II outlines the system model used in this paper and describes the limitations of the LPE criterion. The applicable conditions of the existing LPE criterion, including the consideration of the load and the influence of the cutset selection, are presented in Section III. Based on the analysis of the applicable conditions and limitations of the criterion, two simple improvement strategies are presented in section IV. In section V, the verification of the results with a 39-bus test system is described. Section VI concludes the paper.

## II. SYSTEM MODEL AND LINE POTENTIAL ENERGY CRITERION

### A. System model and its energy function

In this study, the power system consists of $m$ generators and $N$ buses. The basic equations are described below. The models of the generators are not specified because Eq. (3) can be modified according to the required generator model, such as two-order or larger-order models.

$$\frac{d\delta_i}{dt} = \omega_i \quad (1)$$

$$M_i \frac{d\omega_i}{dt} = P_{mi} - P_{ei} - \frac{M_i}{M_T} P_{COI} \quad (2)$$

$$P_{ei} = f(\delta_i, \phi_{Gi}) \quad (3)$$
$$(i=1,2,\ldots m)$$

where,

$$P_{COI} = \sum_{i=1}^{m}(P_{mi} - P_{ei}), M_T = \sum_{i=1}^{m} M_i.$$

The power system with $m$ generators, $ns$ network lines, and $nl$ load nodes $n= m+ns+nl$ is illustrated in Fig. 1. For the convenience of description, the first to $m$-th branches are generator branches, the $(m+1)$-th to $(m+ns)$-th branches are network branches, and the $(m+ns+1)$-th to $(m+ns+nl)$-th branches are load nodes connected to the ground. In a lossless network, the energy function of a multi-machine system is defined as [17]:

$$V = V_{KE} + V_{PE} = \frac{1}{2}\sum_{i=1}^{m} M_i \omega_i^2 + \sum_{i=1}^{n} \int_{\sigma_i^s}^{\sigma_i} [P_i(\theta_i) - P_i^s]d\theta_i \quad (4)$$

where $\sigma_i = \phi_k - \phi_l$ when $i$ denotes network lines; $\sigma_i = \delta_i - \phi_{Gi}$ when $i$ denotes generators. $\phi_{Gi}$ denotes the phase angle of the $i$-th generator terminal bus voltage in the center of the inertia reference. $\sigma_i = \phi_{Li}$ when $i$ denotes load nodes. $\phi_{Li}$ denotes the phase angle of the $i$-th load node voltage in the center of inertia reference. $\sigma_i^s$ means $\sigma_i$ at the stable equilibrium point after a failure. $\theta_i$ is the integration variable and has the same meaning as $\sigma_i$.

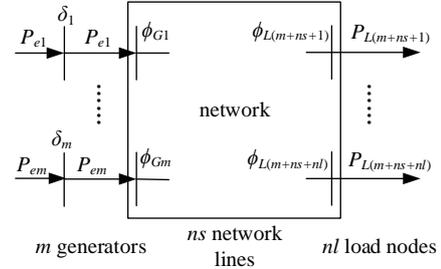

Fig. 1 Simplified diagram of the electrical components in the system

The contribution of each element in $V_{PE}$ is called the LPE. It should be noted that all components in the system, including the load elements, are included in $V_{PE}$.

### B. Line potential energy criterion and its limitations

The LPE criterion is established based on the LPE of the cutset lines to indicate the output of the generators. It has different forms of expression [17,19,20], such as Eqs. (5), (6), and (7), which are derived from the same principle.

$$V_{KE} + \sum_{i \in C} \int_{\sigma_i^s}^{\sigma_i(0)} (P_i - P_i^s) d\theta_i \geq v_{i0}^+, \quad (5)$$

$$\begin{cases} P_k(t_{bk}) - P_k^s = 0, k \in C \\ dP_k(t_{bk})/dt \neq 0 \end{cases} \quad (6)$$

$$\begin{cases} P_i(t) - P_i^s \leq 0, \forall i \in C \\ \sigma_i(t) > \sigma_{\min}, \forall i \in C \\ |d\sigma_i(\xi)/dt| > 0, \forall i \in C, \xi \in [t_0,t] \end{cases} \quad (7)$$

where $C$ denotes the set of cutset lines whose reference direction is from the leading group to the lagging group, $\sigma_i^s$ denotes a stable post-fault equilibrium point, $\sigma_i(0)$ denotes the state after the fault clearance, and $v_{i0}^+$ denotes the cutset vulnerability index corresponding to the selected cutset and the unstable equilibrium point (UEP) [16,17]. $t_{bk}$ denotes the time when the LPE of the $k$-th line reaches its first maximum after fault clearance [19]. $\sigma_{\min}$ is the threshold [20].

The LPE criterion is derived from the structure-preserving energy function. It is assumed that the LPE of the cutset lines determines the instability of the generator rotor angle. This reasoning process is illustrated by the following four steps, which are common in the existing literature:

*Step 1*: The rotor angle instability of the system can be derived from the energy function analysis, e.g., the potential energy boundary surface method. The total energy of the system is expressed as the sum of the kinetic energy and potential energy, as shown in Eq. (4).

*Step 2*: In the energy function of the structure-preserving model, the potential energy of the generator can be expressed as the sum of the LPE in the series elements (transmission line, transformer, and generator reactance) [23].

*Step 3*: Under certain assumptions, the LPE of the series elements can be expressed as the LPE of the "critical cutset", which is also known as "saturated lines" [17,24]. This refers to the state in which the angle differences go unbounded because the LPE of the other series elements is small enough to ignore.

*Step 4*: Because of the previous steps 1-3, the rotor angle instability of the system can be determined according to the LPE of the cutsets.

However, because of the limitations of the LPE reasoning process, the applicability of the criterion has two limitations.

*Limitation 1*: the effect of the load is ignored because of the assumption of a constant power load.

In step 2, as shown in Eq. (4), the total VPE includes not only the series elements but also the parallel elements (the load and any other power injection equipment, except for the generators). If the parallel elements do not have constant active power, the potential energy of the parallel branch may not be small enough to be ignored; thus, it is necessary to determine the LPE of the parallel elements.

*Limitation 2*: the criterion overemphasizes the "critical cutsets".

In step 3, the definition of "critical cutsets" is not clear. A paper was devoted to this issue [18]; moreover, the critical cutsets may drift [25,26]. If the cutset selection is very flexible, or the cutset is selected based on the existing PMU measurement, the LPE criterion is easier to apply to online applications. Another problem is that the online identification of the critical cutset may lead to a delay in estimating the transient stability [27]. The main reason is that the change in the electric variable of the cutset lines does not occur simultaneously, and some of the phase angle changes of the lines may occur more slowly. This leads to a delay in the detection of the critical cutset until all critical cutset lines predict instabilities. The goal of the LPE criterion is to monitor the generators rather than the lines in the network. The analysis results of this study also show that the criterion based on the LPE does not require monitoring the critical cutset in theory.

## III. THE ANALYSIS OF THE LINE POTENTIAL ENERGY CRITERION

In this section, limitations 1 and 2 are analyzed from the perspective of the two-machine equivalent. The LPE characteristics of the constant impedance load are analyzed. Subsequently, the condition of choosing cutsets in this instability criterion is analyzed.

### A. Analysis from the perspective of the two-machine equivalent

It is well-known that not all generator kinetic energy is highly correlated with rotor angle instability. Because the rotor angle instability often occurs in clusters, only the relative kinetic energy between the clusters can contribute to the out-of-step condition [28]. The system is represented by two clusters, which are denoted as network $A$ and network $S$, respectively; these are connected by the cutset lines, as shown in Fig. 2.

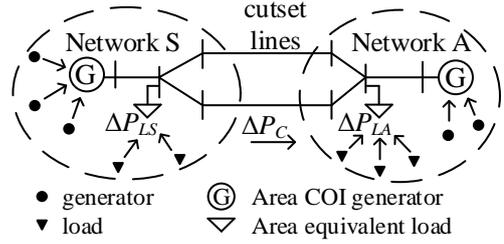

- generator    Ⓖ Area COI generator
- ▼ load    ▽ Area equivalent load

Fig. 2 Simplified schematic diagram of a power grid. The power grid is divided into network $S$ and network $A$ according to the cutset lines. In each network, the generators in the network are equivalent to an inertia center generator, and the loads are regarded as an equivalent load.

Assuming that the rotor angle of the generator in network S is larger than that in network A, the relative kinetic energy between the two clusters are defined as [28]:

$$V_{KE2} = \frac{1}{2}\frac{M_S M_A}{M_S + M_A}\omega_2^2 \quad (8)$$

where

$$\omega_2 = \omega_S - \omega_A, \quad M_S = \sum_{i \in S} M_i, \quad M_A = \sum_{i \in A} M_i,$$

$$\omega_S = \sum_{i \in S} M_i \omega_i / M_S, \quad \omega_A = \sum_{i \in A} M_i \omega_i / M_A.$$

Similar to Eq. (4), according to the relative kinetic energy $V_{KE2}$, the relative LPE $V_{PE2}$ is defined, satisfying Eq. (9). The detailed derivation can be found in Appendix A.

$$V_{KE2}(t_0,t_1) + V_{PE2}(t_0,t_1) = 0 \quad (9)$$

where

$$V_{PE2} = V_{PEC2} + V_{PEL2} + V_{PEloss2},$$

$$V_{PEC2} = \int_{t_0}^{t_1} \Delta P_C \omega_2 dt, \quad V_{PEL2} = \int_{t_0}^{t_1} \Delta P_{L2} \omega_2 dt,$$

$$V_{PEloss2} = \int_{t_0}^{t_1} (\Delta P_{loss2} + \Delta P_{Closs}) \omega_2 dt,$$

$$\Delta P_{L2} = (\Delta P_{LS} M_A - \Delta P_{LA} M_S)/(M_S + M_A),$$

$$\Delta P_{LS} = \sum_{\substack{i=m+ns+1 \\ i \in S}}^{m+ns+nl} P_{Li} - P_{Li}^s, \quad \Delta P_{LA} = \sum_{\substack{i=m+ns+1 \\ i \in A}}^{m+ns+nl} P_{Li} - P_{Li}^s,$$

$$\Delta P_C = \sum_{k \in S}\sum_{l \in A} P_{kl} - P_{kl}^s.$$

and $\Delta P_{loss2}$ represents the equivalent power loss inside the clusters. $\Delta P_{Closs}$ represents the equivalent power loss between the two clusters. $\Delta P_{LS}$ and $\Delta P_{LA}$ denote the equivalent load power in network $S$ and network $A$, respectively, as shown in Fig. 2.

Based on the definition of $V_{PE2}$ in Eq. (9), some results about kinetic energy, such as the stable boundary, can be transformed into the expression of the relative LPE of the lines and loads in the network. This approach makes it easier to understand the application conditions of the LPE instability criterion.

According to the potential energy boundary surface method, when the trajectory crosses the stable boundary, the potential energy reaches the maximum, but the kinetic is not equal to 0, which is expressed as follows.

$$dV_{PE2}/dt = 0 \quad (10)$$

$$d^2V_{PE2}/dt^2 < 0 \quad (11)$$

$$\omega_2 \neq 0 \quad (12)$$

As defined in Eq. (9), the critical state satisfies the following:

$$\Delta P_{L2} + \Delta P_C + \Delta P_{loss2} + \Delta P_{Closs} = 0 \quad (13)$$

$$\frac{d(\Delta P_{L2} + \Delta P_C + \Delta P_{loss2} + \Delta P_{Closs})}{dt}\omega_2 < 0 \quad (14)$$

Equation (13) shows that the power of the cutset lines and the equivalent load power determine the rotor angle instability of the generator based on the two-machine equivalent.

Since the line conductance is far smaller than the load conductance, $\Delta P_{loss2}$ and $\Delta P_{Closs}$ are much smaller than $\Delta P_{L2}$ in Eqs. (13) and (14).

According to Eq. (13), the instability criterion of LPE could be defined as shown in Eq. (15) if the line loss is ignored, which is very similar to Eqs. (6) and (7). The remainder of this paper describes the LPE criterion in the form of Eq. (15) because it can be regarded as a necessary condition of Eq. (7), and the system instability is detected slightly earlier than with Eq. (7). Besides, using this form is more convenient for the analysis based on the two-machine equivalence in theory.

$$\begin{cases} \Delta P_{L2} + \Delta P_C = 0 \\ \dfrac{d(\Delta P_{L2} + \Delta P_C)}{dt}\omega_2 < 0 \end{cases} \quad (15)$$

### B. The Characteristic of The Load Energy

In this subsection, we discuss the effect of the relative LPE of the load on the rotor angle stability of the generator according to the two-machine equivalent.

All buses except for the generators and loads of the network are eliminated. The equation of the bus voltage and injection current in the system is expressed as follows:

$$\begin{pmatrix} -\mathbf{I_L} \\ \mathbf{I_G} \end{pmatrix} = \begin{pmatrix} \mathbf{Y_{LL}} & \mathbf{Y_{LG}} \\ \mathbf{Y_{GL}} & \mathbf{Y_{GG}} \end{pmatrix} \begin{pmatrix} \mathbf{V_L} \\ \mathbf{E_G} \end{pmatrix} \quad (16)$$

where $\mathbf{I_L}$ and $\mathbf{I_G}$ denote the current vector of the load and generators, respectively. $\mathbf{V_L}$ denotes the bus voltage vector of the load node, and $\mathbf{E_G}$ denotes the internal electromotive force vector of the generators.

A constant impedance load is assumed, i.e., $\mathbf{I_L} = [\mathbf{Y_L}]\mathbf{V_L}$, where $[\mathbf{Y_L}] = \text{diag}(Y_{Li})$ denotes the diagonal matrix whose diagonal line is composed of the load admittance $Y_L$. Thus, the bus voltage vector $\mathbf{V_L}$ can be rewritten as Eq. (17).

$$\mathbf{V_L} = \mathbf{Z_{LL}} \mathbf{Y_{LG}} \mathbf{E_G} \quad (17)$$

where $E_{Gi} = |E_{Gi}|\angle \delta_i$ and $\mathbf{Z_{LL}} = -([\mathbf{Y_L}] + \mathbf{Y_{LL}})^{-1}$.

According to the division of the cluster $S$ and cluster $A$, the $\mathbf{E_G}$ can be divided into two vectors, $\mathbf{E_{GS}}$ and $\mathbf{E_{GA}}$, and $\mathbf{V_L}$ can also be divided into two vectors, $\mathbf{V_{LS}}$ and $\mathbf{V_{LA}}$. Thus, Eq. (17) can be rewritten in the following form:

$$\begin{pmatrix} \mathbf{V_{LS}} \\ \mathbf{V_{LA}} \end{pmatrix} = \begin{pmatrix} \mathbf{Z_{SS}} & \mathbf{Z_{SA}} \\ \mathbf{Z_{AS}} & \mathbf{Z_{AA}} \end{pmatrix} \begin{pmatrix} \mathbf{Y_{LGS}} & 0 \\ 0 & \mathbf{Y_{LGA}} \end{pmatrix} \begin{pmatrix} \mathbf{E_{GS}} \\ \mathbf{E_{GA}} \end{pmatrix} \quad (18)$$

Assuming an out-of-step phenomenon of the two clusters,

$$\mathbf{E_{GS}} = \mathbf{E_{GS}^s} \exp(j(\delta_S - \delta_S^s)), \quad (19)$$

$$\mathbf{E_{GA}} = \mathbf{E_{GA}^s} \exp(j(\delta_A - \delta_A^s)), \quad (20)$$

where superscript $s$ denotes the variable at a steady time,

$$\delta_S = \sum_{i \in S} M_i \delta_i \bigg/ M_S, \ \delta_A = \sum_{i \in A} M_i \delta_i \bigg/ M_A.$$

Because of Eqs. (18), (19), and (20), every bus voltage is a function of $\delta_S$ and $\delta_A$. Therefore, $\Delta P_{LS}$, $\Delta P_{LA}$, and $\Delta P_C$ in Eq. (9) can be rewritten as the following equations:

$$\Delta P_{LS} = P_{LSc} + P_{LS\max} \cos(\delta + \gamma_{LS}), \quad (21)$$

$$\Delta P_{LA} = P_{LAc} + P_{LA\max} \cos(\delta + \gamma_{LA}), \quad (22)$$

$$\Delta P_C = P_{Cc} + P_{C\max} \cos(\delta + \gamma_C), \quad (23)$$

where $P_{LSc}$, $P_{LS\max}$, $\gamma_{LS}$, $P_{LAc}$, $P_{LA\max}$, $\gamma_{LA}$, $P_{Cc}$, $P_{C\max}$, $\gamma_C$, are constants defined in Appendix B for a constant $|E_{Gi}|$, and $\delta = (\delta_S - \delta_S^s) - (\delta_A - \delta_A^s)$.

Usually, the $\gamma_{LS}$ is close to $\gamma_{LA}$. Thus, the form of $\Delta P_{L2}$ in Eq. (9) is simplified as follows:

$$\Delta P_{L2} \approx P_{L2c} + P_{L2\max} \cos(\delta + \gamma_{L2}) \quad (24)$$

where

$$P_{L2c} = (P_{LSc} M_A - P_{LAc} M_S)/(M_S + M_A),$$

$$P_{L2\max} = \frac{P_{LS\max} M_A - P_{LA\max} M_S}{M_S + M_A} \cos(\frac{\gamma_{LS} - \gamma_{LA}}{2}),$$

$$\gamma_{L2} = (\gamma_{LS} + \gamma_{LA})/2.$$

The above parameters and variables have the following properties. $\delta > 0$ indicates that the generators in network $S$ are ahead of those in network $A$. $\gamma_{LS}$ and $\gamma_{LA}$ are around 0, and $\gamma_C$ is around $-\pi/2$. The analysis process is shown at the end of Appendix B. The inertia of the generators in network $S$ are usually smaller than those in network $A$, $M_A \gg M_S$. In addition, since $P_{LS\max} > 0$ and $P_{LA\max} > 0$, $P_{L2\max} > 0$ typically applies. Besides, when the steady-state equilibrium point of the system is stable with a small disturbance, $df(0)/d\delta > 0$ is met, which is similar to the synchronizing power coefficient [29]. A brief validation process is shown in Appendix C. Therefore, let

$$f(\delta) = \Delta P_{L2}(\delta) + \Delta P_C(\delta) \quad (25)$$

*Proposition 1*: There exists at least one point of $\delta$ in the interval between 0 and $-2\gamma_C$ that satisfies $f(\delta) = 0$ when the conditions are true: $\delta > 0$, $\gamma_C \in (-\pi + |\gamma_{L2}|, 0)$, $df(0)/d\delta > 0$ and $P_{L2\max} > 0$.

The details on proposition 1 are provided in Appendix D. According to proposition 1, monitoring $\Delta P_C = 0$ (when $\delta = -2\gamma_C$) can replace monitoring $\Delta P_{L2} + \Delta P_C = 0$. This is the reason that the instability criterion of the LPE is theoretically feasible even if the load power is ignored.

Monitoring $\Delta P_C=0$ requires only some local measurements. However, as shown in Fig. 3, in principle, the time of monitoring the system's transient instability occurs later than the direct monitoring method of the generator rotor angle, although the prediction method based on curve fitting is used in reference [20] to improve the timeliness of this monitoring method. In section IV, a method is proposed to compensate for this shortcoming. However, the curve prediction method can also be used.

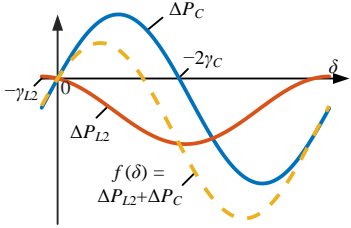

Fig. 3 Diagrams of $\Delta P_C$, $\Delta P_{L2}$, and $\Delta P_C + \Delta P_{L2}$.

If there is no voltage collapse caused by the constant power load, such as singular induced bifurcation, the constant power load does not contribute to the relative LPE in Eq. (9).

However, in the transient process, the program logic in many simulation programs converts the constant power load to impedance when the voltage is lower than a certain value [30] because it is difficult to maintain the load power in a real scenario when the voltage is very low. Examples of simulation programs include PSS/E [31], fix load in PSCAD/EMTDC [29], and the Three-Phase Dynamic Load in SIMULINK. Thus, the characteristics of the constant power are similar to those of the constant impedance load, as shown in Fig. 3.

In summary, proposition 1 is also suitable for a system with a constant power load.

### C. THE SELECTION OF THE CUTSET FOR OUT-OF-STEP DETECTION

Due to limitation 2, the critical cutset needs to be selected. Although several methods exist to select critical cutsets (or vulnerable cutsets), it is proved in this subsection that the cutset for out-of-step detection is not limited to critical cutsets.

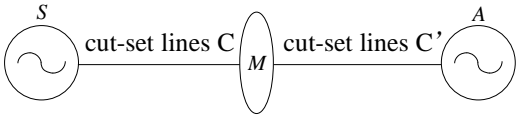

Fig. 4 Two different cutset lines ($C$ and $C'$) with the same generator cluster division.

As defined in Eq. (9), the cutset lines $C$ and $C'$, which have the same division, provide the same $V_{KE2}$ and $V_{PE2}$. The two different cutset lines provide different $V_{PEC2}$ and $V_{PEL2}$ (and different $V_{PEloss2}$), but $V_{PEC2}+V_{PEL2}$ remains nearly constant (because $V_{PEloss2}$ can be ignored compared with $V_{PEL2}$). The reason is that $\Delta P_C+\Delta P_{L2}$ remains nearly constant.

$\Delta P_{L2}$ and $\Delta P_C$ refer to the cutset lines $C$. $\Delta P_{L2}{\sim}$ and $\Delta P_C{\sim}$ refer to the cutset lines $C'$. Thus, $\Delta P_{L2}{\sim}$ and $\Delta P_C{\sim}$ have the following relationship with $\Delta P_{L2}$ and $\Delta P_C$:

$$\Delta \tilde{P}_{L2} = \frac{(\Delta P_S + \Delta P)M_A - (\Delta P_A - \Delta P)M_S}{M_A + M_S} \quad (26)$$
$$= \Delta P_{L2} + \Delta P$$

$$\Delta \tilde{P}_C \approx \Delta P_C - \Delta P \quad (27)$$

where

$$\Delta P = \sum_{i \in M} P_{Li} - P_{Li}^s$$

Therefore, $\Delta P_C{\sim}+\Delta P_{L2}{\sim}\approx \Delta P_C+\Delta P_{L2}$. As mentioned regarding the effect of the load energy, as long as the condition in Proposition 1 is met, Eq. (7) is valid to detect the rotor angle stability. Therefore, cutset C', which has the same instability mode as cutset C, provides the same result of the generator instability. This finding shows that, for the same instability mode, the selection of the cutsets does not affect the result of the generator out-of-step detection.

For ordinary cutsets, since the LPE of the cutset lines is not higher than that of the other lines, this result seems to contradict the requirements of step 3 in section II. The main reason is that the LPE of the cutset lines is different from the relative LPE $V_{PE2}$. The potential energy in the derivation process in this section is the relative LPE $V_{PE2}$ rather than the LPE of cutset lines. Besides, the angle velocity difference $\omega_2$ is not determined by the cutset lines. Therefore, using the relative LPE $V_{PE2}$ changes the condition for selecting the cutset lines.

This finding also shows that selecting certain cutset lines results in a smaller $\Delta P_{L2}$, leading to an earlier instability detection by only using $\Delta P_C=0$.

### D. THE MODIFIED REASONING LOGIC

Based on the analysis of limitation 1 and limitation 2, the reasoning logic of the online detection can be improved as follows:

*Step 1*: The rotor angle stability of the system can be determined by analyzing the relative energy of the cluster, like the potential energy boundary surface method.

*Step 2*: The relative energy of the generator can be expressed as the sum of the relative LPE in the cutset, the load elements, and all network line losses.

*Step 3*: Given the proposed assumptions, the relative LPE in the cutset is related to that in the load elements; the power of the cutset lags behind the load power by around π/2 in phase. According to proposition 1, the relative LPE in the cutset can be used to infer the maximum value of the relative LPE.

*Step 4*: Since steps 1-3 are used, the rotor angle instability of the system can be determined according to the relative LPE of the cutsets.

After the theoretical analysis, it was concluded that limitation 1 and limitation 2 do not represent limitations of the LPE criterion for the online detection of the generator out-of-step. Instead, the reasoning logic led to these limitations.

## IV. IMPROVED LINE POTENTIAL ENERGY CRITERION

The previous section theoretically analyzed the applicable conditions of the LPE, which provided support for its online application. Based on the above analysis, this section proposes two simple and effective strategies to improve the LPE criterion. These two strategies can be used at the same time or separately in appropriate situations.

*A. A STRATEGY BASED ON LOAD POWER*

The above analysis shows that the contribution of the load power changes to the generator out-of-step condition is not necessarily detected by power line monitoring. Therefore, this section discusses a simple method to detect the system transient stability earlier by monitoring a few loads.

According to Eq. (9), since $M_A \gg M_S$, $\Delta P_{LS}$ is dominant in $\Delta P_{L2}$, and only loads in Network S have to be monitored.

Because of Eq. (18), if the monitored load bus is set $N$, $N \in S$. Then the load power in set $N$ can be written as Eq. (28) similar to Eq. (21).

$$\Delta P_{LS,N} = P_{LSc,N} + P_{LS\max,N} \cos(\delta + \gamma_{LS,N}), \quad (28)$$

where $P_{LSc,N}$, $P_{Smax,N}$, and $\gamma_{LS,N}$ are constant. Their expressions can be directly derived by setting the load conductance $G_{Li}$, except for the load bus in set $N$, to zero in Eq. (21) or the expression (B2) in the Appendix B.

When $\gamma_{LS,N} \approx \gamma_{LS}$, The variation of equivalent load power in area $S$ can be approximated in the following way:

$$\Delta P_{LS} \approx \alpha \Delta P_{LS,N} = \alpha \sum_{i \in N}(P_{Li} - P_{Li}^s) \quad (29)$$

where,

$\alpha = \min(\alpha_1, \alpha_2)$,

$\alpha_1 = P_{LSc,N}/P_{LSc}$, $\alpha_2 = P_{LS\max,N}/P_{LS\max}$.

Since $\alpha_1$ and $\alpha_2$ are relatively close, the smaller value of the two is used to ensure the reliability of the result. Therefore, Eq. (29) is substituted into (9), omit $\Delta P_{LA}$, and obtain the estimated value of $\Delta P_{L2}$, as shown in Eq. (30). Equation (13) can be used to determine if the system is unstable.

$$\Delta P_{L2} \approx \alpha M_A/(M_A + M_S)\sum_{i \in N}(P_{Li} - P_{Li}^s) \quad (30)$$

*B. A STRATEGY BASED ON CUTSET SELECTION*

The choice of the cutset position does not affect the evaluation of the system instability by the relative LPE, but it does will affect the evaluation time because of the relationship between the cutset power and load power, as shown in Fig. 3. Therefore, when there are multiple cutsets available to monitor the generator instability, selecting certain cutsets may result in a faster evaluation of the system instability.

According to the analysis of Eq. (24), due to the difference in the generator inertia between cluster $S$ and cluster $A$, $P_{L2\max}>0$ is reasonable, i.e., $\Delta P_{L2}$ is typically positive. Besides, the bus voltage in the system decreases during the swing-out process of the generator. Therefore, according to Eq. (26), $\Delta P_{L2}$ obtained by the cutset $C$ close to cluster $S$ is closer to 0 than $\Delta P_{L2}\tilde{}$ obtained by the cutset C' close to cluster $A$. As shown in Fig. 5, Point $F$ ($\Delta P_C=0$) is closer to Point $D$ ($\Delta P_C+\Delta P_{L2}=0$ the stability boundary is approximately defined by Eq. (13) if the line power loss is ignored) than Point $F'$ ($\Delta P_C\tilde{}=0$). Thus, if only the instability of the cutset power is monitored, Point $F$ of cutset C is detected earlier than Point $F'$ of cutset C'.

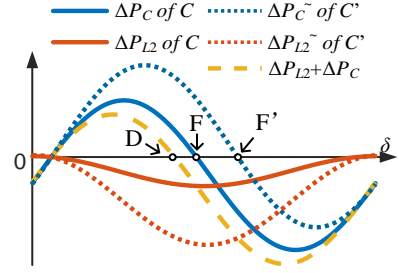

Fig. 5 Comparison of $\Delta P_C$ and $\Delta P_{L2}$ in different cutsets.

In short, when there are multiple cutsets with the same generator grouping mode, the cutset close to the generator group with the lower inertia can detect the instability of the generator group earlier. One extreme case occurs when a single generator is out-of-step with the remaining generators; it may have better performance to monitor the generator instability according to the cutset lines of the LPE at the generator outlet [32,33].

V. VERIFICATION

In this section, the results of the theoretical analysis are verified by the 39-bus New England system, including the characteristics of the relative LPE of the load (proposition 1) and the performance of the cutset criteria for different cutsets.

The single-line diagram of the New England system is shown in Fig. 6. Two instability modes in different steady-state operations are tested. 1) Generator 39 lags behind the other generators, and 2) generators 33, 34, 35, and 36 are ahead of the other generators. In instability mode 1, set $C_1$ consisting of lines 1-2 and 8-9 is the critical cutset that separates the system into two areas, as shown in Fig. 7 (a). $C_2$ (consisting of lines 1-39 and 9-39) and $C_3$ (consisting of lines 1-39, 17-18, 4-14, 5-6, and 6-7) are two other cutsets. In instability mode 2, set $C_4$ is the critical cutset consisting of lines 16-17 and 15-16, as shown in Fig. 7 (b). $C_5$ consisting of lines 21-22, 23-24, and 16-19 is the other cutset.

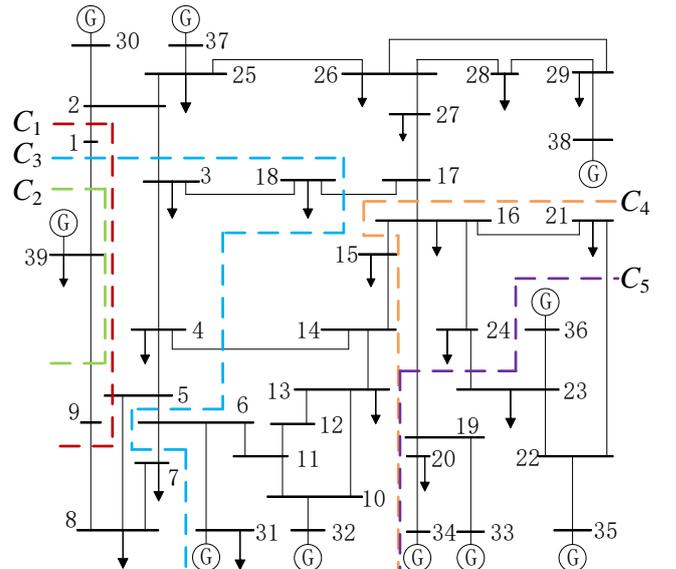

Fig. 6 Single-line diagram of the 10 generators in the New England system

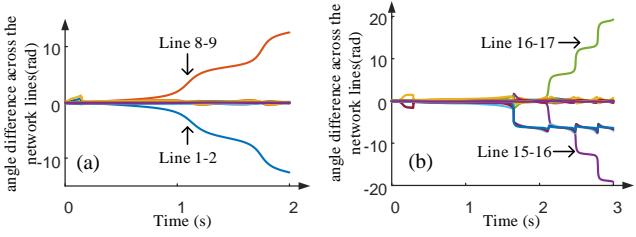

Fig. 7 Angle difference in the network lines: (a) fault on bus 24 with a fault duration of 0.14 s; (b) fault on bus 3 with a fault duration of 0.29 s.

### A. Verification of the Load Energy Characteristics

Set $C_1$ is used as an example, and the derivation results of Eqs. (23) and (24) for the load power are verified by the simulation, as shown in Fig. 8. The curve shows that the simplified equivalent load power approximates the simulation curve.

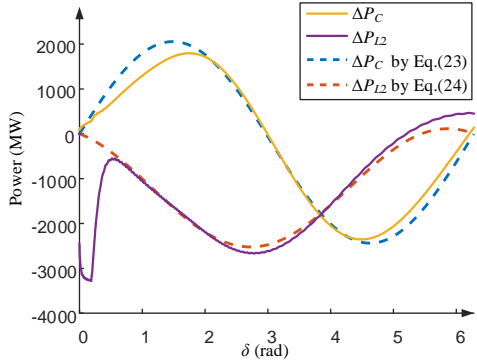

Fig. 8 Comparison of calculation results derived from the Equations and the simulation

Figure. 9 compares the curves of $\Delta P_C$ and $\Delta P_{L2}$ of cutsets $C_1$, $C_2$, and $C_3$. At the position of $\delta=2.783$ rad, the $\Delta P_{L2}$ of $C_1$ reaches the minimum value, i.e., at 180° of the cosine; thus, the estimate of $\gamma_{L2}$ is 0.3586 rad ($\approx$ 20.55°). The $\Delta P_C$ of $C_1$ reaches the first minimum at the position of $\delta=4.478$ rad; thus, the estimate of $\gamma_C$ is -1.3364 rad ($\approx$ -76.57°). Therefore, $\Delta P_C$ of $C_1$ lags behind $P_{L2}$ of $C_1$ by 1.695 rad ($\approx$ 97.11°), $P_{L2\max} = 12.6757 > 0$, which satisfies the condition of proposition 1. The curves $\Delta P_C$ and $\Delta P_{L2}$ of $C_2$ and $C_3$ provide similar results.

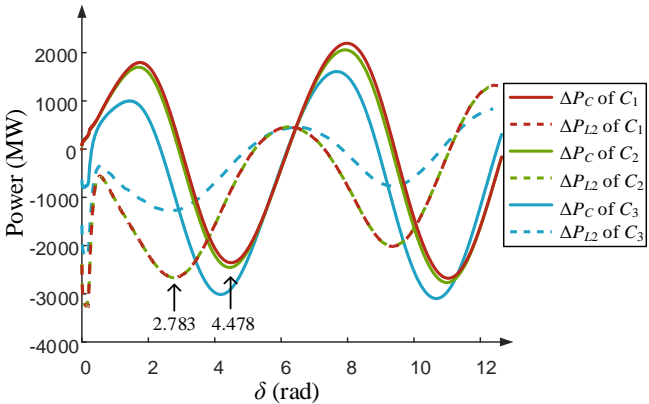

Fig. 9 Relationship between the power of the load and the power of the cutset in the out-of-step condition (fault on bus 24)

According to proposition 1, $\Delta P_C=0$ can be used to detect the loss of synchronization. In Fig. 10 (a), Point $D$ ($D_1$, $D_2$, and $D_3$) corresponds to the point defined in Eq. (13), and point $B$ corresponds to the maximum potential energy, i.e., the point defined in Eq. (10). Because of the effect of the network line loss, the points $B$ and $D$ do not occur at the same time. Point $F$ ($F_1$, $F_2$, and $F_3$) corresponds to the point defined in Eq. (15). Thus, Point $D$ ($D_1$, $D_2$, and $D_3$) and $B$ can be approximately seen as the boundary of the stability, and $F$ ($F_1$, $F_2$, and $F_3$) is used to detect whether the system is unstable. The curves of instability mode 2 are similar to those of instability mode 1.

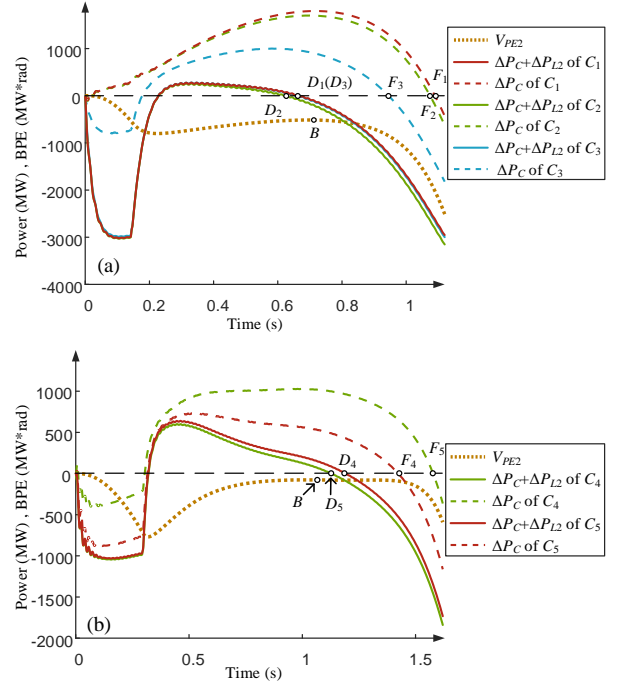

Fig. 10 Influence of the load power on the LPE criterion: (a) instability mode 1; (b) instability mode 2

The time when point $F$ ($F_1$, $F_2$ and $F_3$) detects the instability is later than when Point $D$ ($D_1$, $D_2$ and $D_3$) detects it. As proposed in the previous section, monitoring parts of the load power can reduce this delay. For cutset $C_1$, the influence of the load energy on the generator instability is estimated by monitoring the load power of bus 3, bus 4, bus 7, and bus 8 in the network $S$. The parameters $P_{LSc,N}$, $P_{Smax,N}$, and $\gamma_{LS,N}$ in Eq. (28) are calculated. Since $\gamma_{LS,N}$ (=0.3408 rad) is close to $\gamma_{L2}$(=0.3586 rad), the condition of Eq. (29) is met. Then the parameter $\alpha$ is calculated using Eq. (29); $\alpha_1$=2.1798, $\alpha_2$=2.2611, and $\alpha$=2.1798. Finally, $\Delta P_{L2}$ is estimated using Eq. (30). For cutset $C_4$, bus 21 is used as an example of load power monitoring. Similarly, the following parameters are obtained: $\gamma_{LS,N}$=0.2729 rad ($\approx \gamma_{L2}$=0.2764 rad), $\alpha_1$= 5.8557, $\alpha_2$= 5.8655, and $\alpha$=5.8655. The purple dotted line in Fig. 11 shows the estimation results, demonstrating that this method accelerates instability detection. The proposed method predicts the instability more than 300 ms earlier (at Point $D_1$' compared to Point $F_1$) and more than 380 ms earlier (at $D_4$' compared to $F_4$).

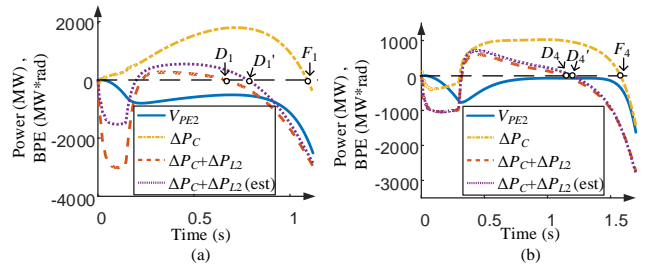

Fig. 11 Estimation of the $\Delta P_{L2}$ component using partial load measurements: (a) cutset $C_1$ using the load power of bus, 3, 4, 7, 8; (b) cutset $C_4$ using the load power of bus 21.

### B. Verifying the LPE Criterion for Different Cutsets

Three different cutsets $C_1$, $C_2$, and $C_3$ are selected to compare the influence of different cutsets on the LPE criterion. As seen in Fig. 12, the largest proportion of the sum of the LPE of all the series elements is the LPE of $C_1$, which is the critical cutset. In contrast, the LPE of the other two cutsets is very small, indicating that $C_2$ and $C_3$ do not meet the condition of step 3 in section II.

However, Fig. 9 shows that all three cutsets satisfy the condition of proposition 1. Therefore, the LPE criterion of any of the three cutsets is valid. It can also be verified from Fig. 10 that Points $F_1$, $F_2$, and $F_3$ provide similar results for detecting the rotor angle instability.

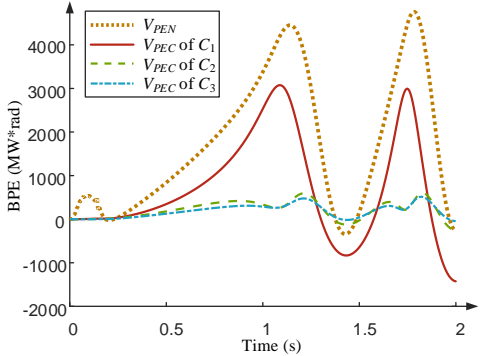

Fig. 12 LPE of different cutsets $C_1$, $C_2$, and $C_3$. $V_{PEN}$ denotes the sum of the LPE of all the series elements, and $V_{PEC}$ denotes the LPE of the cutset.

Another method to improve the performance of the indicators is through cutset selection. In Fig. 10 (a), the cutset $C_3$ detects the system instability more than 140 ms earlier than $C_1$ and $C_2$, but the cutset $C_3$ contains 5 lines, which is inconvenient. However, in the other cases, a reasonable choice of the cutset assembly provides good results, for example, choosing cutset $C_5$ instead of cutset $C_4$, If only the power of the cutset lines is used, $C_5$ detects the instability more than 150 ms earlier than $C_4$, as shown in Fig. 10 (b). This finding verifies the previous results, i.e., choosing a cutset closer to the low-inertia generator group results in earlier detection of the generator instability.

## VI. CONCLUSION

In this paper, we discussed the limitations of the LPE criterion and analyzed the characteristics of the criterion to extend its application. The effect of the load in step 2 was discussed, and the selection condition of the cutset in step 3 was relaxed.

The theoretical analysis showed that the load did not affect the reliability of the LPE criterion but increased the delay of the LPE criterion. The position of the cutset did not affect the condition of the LPE criterion but did affect the detection time of the LPE criterion. This paper proposed two simple improvements based on the load power compensation and cutset location selection to detect the system instability earlier and increase the effectiveness of the subsequent control.

The proposed analysis can also be used for indices that are similar to the LPE index. In a future study, we will use the LPE criterion to monitor the rotor angle stability of an AC/DC hybrid system to predict unstable conditions.

## APPENDIX A

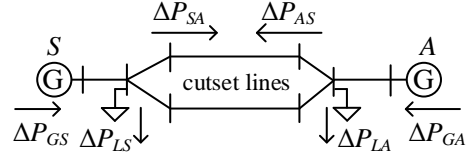

Fig. B1 The two-machine system model

As shown in Fig. B1, Eqs. (A2) and (A1) hold due to the system power balance in networks $A$ and $S$.

$$\Delta P_{GS}+\Delta P_{LS}+\Delta P_{lossS}+\Delta P_{SA}=0 \quad (A1)$$

where

$$\Delta P_{GS}=\sum_{\substack{i=1\\i\in S}}^{m} P_{mi}-P_{ei}, \quad \Delta P_{LS}=\sum_{\substack{i=m+ns+1\\i\in S}}^{m+ns+nl} P_{Li}-P_{Li}^{s},$$

$$\Delta P_{lossS}=\sum_{\substack{i=m+1\\i\in S}}^{m+ns}(P_{i,kl}+P_{i,lk})-(P_{i,kl}^{s}+P_{i,lk}^{s}),$$

$$\Delta P_{SA}=\sum_{k\in S}\sum_{l\in A}P_{kl}-P_{kl}^{s}.$$

Similarly,

$$\Delta P_{GA}+\Delta P_{LA}+\Delta P_{lossA}+\Delta P_{AS}=0 \quad (A2)$$

where

$$\Delta P_{GA}=\sum_{\substack{i=1\\i\in A}}^{m} P_{mi}-P_{ei}, \quad \Delta P_{LA}=\sum_{\substack{i=m+ns+1\\i\in A}}^{m+ns+nl} P_{Li}-P_{Li}^{s},$$

$$\Delta P_{lossA}=\sum_{\substack{i=m+1\\i\in A}}^{m+ns}(P_{i,kl}+P_{i,lk})-(P_{i,kl}^{s}+P_{i,lk}^{s}),$$

$$\Delta P_{AS}=\sum_{k\in A}\sum_{l\in S}P_{kl}-P_{kl}^{s}.$$

Because of Eq. (2), Eqs. (A3) and (A4) are obtained.

$$M_A M_S \frac{d\omega_S}{dt}=M_A(\Delta P_{GS}-\frac{M_S}{M_T}P_{COI}) \quad (A3)$$

$$M_S M_A \frac{d\omega_A}{dt}=M_S(\Delta P_{GA}-\frac{M_A}{M_T}P_{COI}) \quad (A4)$$

Subtracting Eq. (A4) from Eq. (A3), multiplying $\omega_2$ on both sides of the equation, and integrating to obtain $V_{KE2}$ in Eq. (8) yields:

$$V_{KE2}=\int\frac{M_A\Delta P_{GS}-M_S\Delta P_{GA}}{M_A+M_S}\omega_2 dt \quad (A5)$$

$\Delta P_{GA}$ and $\Delta P_{GS}$ are eliminated by integrating Eqs. (A2) and (A1) into Eq. (A5) to obtain the relative potential energy $V_{PE2}$ between the two clusters so that $V_{KE2}+V_{PE2}$ is a constant. The expression of $V_{PE2}$ is:

$$V_{PE2}=\int(\Delta P_{L2}+\Delta P_{loss2}+\Delta P_C+\Delta P_{Closs})\omega_2 dt \quad (A6)$$

where

$$\Delta P_{L2}=(M_A\Delta P_{LS}-M_S\Delta P_{LA})/(M_A+M_S),$$
$$\Delta P_{loss2}=(M_A\Delta P_{lossS}-M_S\Delta P_{lossA})/(M_A+M_S),$$

$$\Delta P_C = \Delta P_{SA},$$

$$\Delta P_{Closs} = -\sum_{k \in S}\sum_{l \in A} \frac{M_S}{M_A + M_S}(P_{kl} + P_{lk} - P_{kl}^s - P_{lk}^s).$$

Therefore,
$$V_{KE2} + V_{PE2} = 0.$$

APPENDIX B

Because of Eq. (B1),
$$\sum_{\substack{i=m+ns+1 \\ i \in S}}^{m+ns+nl} P_{Li} = \mathbf{V}_{LS}^T [\mathbf{G}_{LS}] \mathbf{V}_{LS}^*, \tag{B1}$$

$\Delta P_{LS}$ in Eq. (A1) can be rewritten as:
$$\Delta P_{LS} = \sum_{\substack{i=m+ns+1 \\ i \in S}}^{m+ns+nl} P_{Li} - P_{Li}^s \\ = CstS + real(\mathbf{E}_{GS}^T \mathbf{YS}_{SA} \mathbf{E}_{GA}^*) - \sum P_{LS}^s \tag{B2}$$

where $[\mathbf{G}_{LS}]$ denotes the diagonal matrix consisting of the load conductance of network $S$, and $real(\cdot)$ means the real part of the complex number.

$$CstS = \mathbf{E}_{GS}^T \mathbf{YS}_{SS} \mathbf{E}_{GS}^* + \mathbf{E}_{GA}^T \mathbf{YS}_{AA} \mathbf{E}_{GA}^*,$$
$$\mathbf{YS}_{SS} = \mathbf{Y}_{LGS}^T \mathbf{Z}_{SS}^T [\mathbf{G}_{LS}] \mathbf{Z}_{SS}^* \mathbf{Y}_{LGS}^*,$$
$$\mathbf{YS}_{AA} = \mathbf{Y}_{LGA}^T \mathbf{Z}_{SA}^T [\mathbf{G}_{LS}] \mathbf{Z}_{SA}^* \mathbf{Y}_{LGA}^*,$$
$$\mathbf{YS}_{SA} = 2\mathbf{Y}_{LGS}^T \mathbf{Z}_{SS}^T [\mathbf{G}_{LS}] \mathbf{Z}_{SA}^* \mathbf{Y}_{LGA}^*.$$

Similarly, $\Delta P_{LA}$ in Eq. (A2) can be rewritten as:
$$\Delta P_{LA} = \sum_{\substack{i=m+ns+1 \\ i \in A}}^{m+ns+nl} P_{Li} - P_{Li}^s \\ = CstA + real(\mathbf{E}_{GS}^T \mathbf{YA}_{SA} \mathbf{E}_{GA}^*) - \sum P_{LA}^s \tag{B3}$$

where $[\mathbf{G}_{LA}]$ denotes the diagonal matrix consisting of the load conductance of network $A$ and

$$CstA = \mathbf{E}_{GS}^T \mathbf{YA}_{SS} \mathbf{E}_{GS}^* + \mathbf{E}_{GA}^T \mathbf{YA}_{AA} \mathbf{E}_{GA}^*,$$
$$\mathbf{YA}_{SS} = \mathbf{Y}_{LGS}^T \mathbf{Z}_{AS}^T [\mathbf{G}_{LA}] \mathbf{Z}_{AS}^* \mathbf{Y}_{LGS}^*,$$
$$\mathbf{YA}_{AA} = \mathbf{Y}_{LGA}^T \mathbf{Z}_{AA}^T [\mathbf{G}_{LA}] \mathbf{Z}_{AA}^* \mathbf{Y}_{LGA}^*,$$
$$\mathbf{YA}_{SA} = 2\mathbf{Y}_{LGS}^T \mathbf{Z}_{AS}^T [\mathbf{G}_{LA}] \mathbf{Z}_{AA}^* \mathbf{Y}_{LGA}^*.$$

Then, for the convenience of description, the admittance matrix $\mathbf{Y}_{LL}$ is divided into the following forms in the same manner as $\mathbf{Z}_{LL}$.

$$\mathbf{Y}_{LL} = \begin{pmatrix} \mathbf{Y}_{SS} & \mathbf{Y}_{SA} \\ \mathbf{Y}_{AS} & \mathbf{Y}_{AA} \end{pmatrix}$$

The admittance matrix $\mathbf{Y}_{LL}$ is written as the sum of two parts according to the different branches of the elements that form the admittance matrix. The $\mathbf{Y2}_{LL}$ part is the admittance matrix element consisting of the cutset lines, and the rest is represented as $\mathbf{Y1}_{LL}$. $\mathbf{G1}_S$, $\mathbf{G1}_A$, $\mathbf{G2}_{SS}$, $\mathbf{G2}_{AA}$, $\mathbf{G}_{SA}$, $\mathbf{G}_{AS}$, respectively to represent the real part of the $\mathbf{Y1}_S$, $\mathbf{Y1}_A$, $\mathbf{Y2}_{SS}$, $\mathbf{Y2}_{AA}$, $\mathbf{Y}_{SA}$, and $\mathbf{Y}_{AS}$.

$$\mathbf{Y}_{LL} = \mathbf{Y1}_{LL} + \mathbf{Y2}_{LL} \tag{B4}$$

where
$$\mathbf{Y1}_{LL} = \begin{pmatrix} \mathbf{Y1}_S & 0 \\ 0 & \mathbf{Y1}_A \end{pmatrix}, \mathbf{Y2}_{LL} = \begin{pmatrix} \mathbf{Y2}_{SS} & \mathbf{Y}_{SA} \\ \mathbf{Y}_{AS} & \mathbf{Y2}_{AA} \end{pmatrix}.$$

For the $i$-th cut line, which connects the node $ki$ in cluster $S$ and the node $li$ in cluster $A$, the branch power is denoted as $P_{ki,li}$:
$$P_{ki,li} = V_{ki}^T [G_{ki,li}] V_{ki}^* - real(V_{ki}^T [Y_{ki,li}^*] V_{li}^*). \tag{B5}$$

Thus, $\Delta P_C$ in Eq. (A6) can be rewritten as:
$$\Delta P_C = \sum_{i \in C} P_i - P_i^s \\ = CstC + real(\mathbf{E}_{GS}^T \mathbf{YC}_{SA} \mathbf{E}_{GA}^*) - \sum P_{kl}^s \tag{B6}$$

where
$$CstC = \mathbf{E}_{GS}^T \mathbf{YC1}_{SS} \mathbf{E}_{GS}^* + \mathbf{E}_{GA}^T \mathbf{YC1}_{AA} \mathbf{E}_{GA}^* \\ + real(\mathbf{E}_{GS}^T \mathbf{YC2}_{SS} \mathbf{E}_{GS}^* + \mathbf{E}_{GA}^T \mathbf{YC2}_{AA} \mathbf{E}_{GA}^*),$$
$$\mathbf{YC1}_{SS} = \mathbf{Y}_{LGS}^T \mathbf{Z}_{SS}^T \mathbf{G2}_{SS} \mathbf{Z}_{SS}^* \mathbf{Y}_{LGS}^*,$$
$$\mathbf{YC1}_{AA} = \mathbf{Y}_{LGA}^T \mathbf{Z}_{SA}^T \mathbf{G2}_{SS} \mathbf{Z}_{SA}^* \mathbf{Y}_{LGA}^*,$$
$$\mathbf{YC2}_{SS} = \mathbf{Y}_{LGS}^T \mathbf{Z}_{SS}^T \mathbf{Y}_{SA}^* \mathbf{Z}_{AS}^* \mathbf{Y}_{LGS}^*,$$
$$\mathbf{YC2}_{AA} = \mathbf{Y}_{LGA}^T \mathbf{Z}_{SA}^T \mathbf{Y}_{SA}^* \mathbf{Z}_{AA}^* \mathbf{Y}_{LGA}^*,$$
$$\mathbf{YC}_{SA} = 2\mathbf{Y}_{LGS}^T \mathbf{Z}_{SS}^T \mathbf{G2}_{SS} \mathbf{Z}_{SA}^* \mathbf{Y}_{LGA}^* \\ + \mathbf{Y}_{LGS}^T \mathbf{Z}_{SS}^T \mathbf{Y}_{SA}^* \mathbf{Z}_{AA}^* \mathbf{Y}_{LGA}^* \\ + \mathbf{Y}_{LGS}^T \mathbf{Z}_{AS}^T \mathbf{Y}_{AS} \mathbf{Z}_{SA}^* \mathbf{Y}_{LGA}^*.$$

As an approximate analysis, the dynamics between the generators in the cluster can be neglected, assuming that the generators within the clusters approximately satisfy:
$$\delta_i - \delta_i^s = \delta_S - \delta_S^s, (\forall i \in S).$$

Subsequently:
$$E_{GSi} = |E_{GSi}| \angle (\delta_i^s + \delta_i - \delta_i^s) = E_{GSi}^s \exp(j(\delta_S - \delta_S^s)), \tag{B7}$$

and,
$$\mathbf{E}_{GS} = \mathbf{E}_{GS}^s \exp(j(\delta_S - \delta_S^s)). \tag{B8}$$

Similarly,
$$\mathbf{E}_{GA} = \mathbf{E}_{GA}^s \exp(j(\delta_A - \delta_A^s)), \tag{B9}$$

Thus, the expression of $\Delta P_{LS}$ can be simplified as follows because of Eqs. (B2), (B8) and (B9).
$$\Delta P_{LS} = P_{LSc} + P_{LS\max} \cos(\delta + \gamma_{LS}) \tag{B10}$$

where
$$\delta = (\delta_S - \delta_S^s) - (\delta_A - \delta_A^s),$$
$$P_{LSc} = -real((\mathbf{E}_{GS}^s)^T \mathbf{YS}_{SA} (\mathbf{E}_{GA}^s)^*),$$
$$P_{LS\max} = |(\mathbf{E}_{GS}^s)^T \mathbf{YS}_{SA} (\mathbf{E}_{GA}^s)^*|,$$
$$\gamma_{LS} = angle((\mathbf{E}_{GS}^s)^T \mathbf{YS}_{SA} (\mathbf{E}_{GA}^s)^*),$$

and $angle(\cdot)$ means the angle of the complex number.

Similarly, the expression of $\Delta P_{LA}$ can be simplified as follows:
$$\Delta P_{LA} = P_{LAc} + P_{LA\max} \cos(\delta + \gamma_{LA}), \tag{B11}$$

where
$$P_{LAc} = -real((\mathbf{E}_{GS}^s)^T \mathbf{YA}_{SA} (\mathbf{E}_{GA}^s)^*),$$
$$P_{LA\max} = |(\mathbf{E}_{GS}^s)^T \mathbf{YA}_{SA} (\mathbf{E}_{GA}^s)^*|,$$
$$\gamma_{LA} = angle((\mathbf{E}_{GS}^s)^T \mathbf{YA}_{SA} (\mathbf{E}_{GA}^s)^*).$$

Similarly, the expression of $\Delta P_C$ can be simplified as follows:
$$\Delta P_C = P_{Cc} + P_{SA\max} \cos(\delta + \gamma_C), \tag{B12}$$

where
$$P_{Cc} = -real((\mathbf{E}_{GS}^s)^T \mathbf{YC}_{SA}(\mathbf{E}_{GA}^s)^*),$$
$$P_{LS\max} = |(\mathbf{E}_{GS}^s)^T \mathbf{YC}_{SA}(\mathbf{E}_{GA}^s)^*|,$$
$$\gamma_C = angle((\mathbf{E}_{GS}^s)^T \mathbf{YC}_{SA}(\mathbf{E}_{GA}^s)^*).$$

According to Eqs. (B10) and (B11), $\gamma_S$ and $\gamma_A$ are determined by $\mathbf{YS}_{SA}$ and $\mathbf{YA}_{SA}$, respectively. The absolute values of the imaginary parts of the elements in the matrix $\mathbf{Z}_{LL}$ (consisting of $\mathbf{Z}_{SS}$, $\mathbf{Z}_{SA}$, $\mathbf{Z}_{AS}$, and $\mathbf{Z}_{AA}$) and $\mathbf{Y}_{LG}$ (consisting of $\mathbf{Y}_{LGS}$ and $\mathbf{Y}_{LGA}$) are much larger than the real parts, and $\gamma_S$ and $\gamma_A$ are around 0.

According to Eq. (B12), $\gamma_C$ is determined by $\mathbf{YC}_{SA}$ in (B6). Due to the diagonal dominance of the network impedance components, the absolute values of the elements in $\mathbf{Z}_{SS}$ and $\mathbf{Z}_{AA}$ are significantly greater than those in $\mathbf{Z}_{SA}$ and $\mathbf{Z}_{AS}$. Thus, $\gamma_C$ is more affected by $\mathbf{Y}_{LGS}^T\mathbf{Z}_{SS}^T\mathbf{Y}_{SA}^*\mathbf{Z}_{AA}^*\mathbf{Y}_{LGA}^*$. Since the phase angle of this term is close to $-\pi/2$, $\gamma_C$ is around $-\pi/2$.

APPENDIX C

The generators in network $S$ and network $A$ are equivalent to generator $S$ and generator $A$, respectively, as shown in Fig. B1. Thus, the necessary condition for the steady-state stability of the equivalent two-machine system is $df(\delta)/d\delta>0$.

Due to Eqs. (1) and (2), the rotor motion equation of the equivalent generator can be expressed as follows:
$$\frac{d\delta_l}{dt} = \omega_l, (l=S, A) \tag{C1}$$
$$\frac{d\omega_l}{dt} = \frac{\Delta P_{Gl}}{M_l} - \frac{1}{M_T}P_{COI}, (l=S, A) \tag{C2}$$

By subtracting the rotor equation of equivalent generator $A$ from the rotor equation of the equivalent generator $S$ and using $\delta_2=\delta_S-\delta_A= \delta+\delta_S^s-\delta_A^s$ and $\omega_2=\omega_S-\omega_A$ as the state variables, the linearized model near the equilibrium point is defined as:
$$\begin{bmatrix}\Delta\dot{\delta}_2 \\ \Delta\dot{\omega}_2\end{bmatrix} = \begin{bmatrix}0 & 1 \\ \frac{K_S}{M_S}-\frac{K_A}{M_A} & 0\end{bmatrix}\begin{bmatrix}\Delta\delta_2 \\ \Delta\omega_2\end{bmatrix}, \tag{C3}$$
where
$$K_S = d\Delta P_{GS}/d\delta_2, \; K_S = d\Delta P_{GA}/d\delta_2.$$

The eigenvalue of the coefficient matrix of Eq. (C3) is equal to:
$$\lambda = \pm\sqrt{K_S/M_S - K_A/M_A}. \tag{C4}$$

Eqs. (A1) and (A2) are substituted into Eq. (C3) to eliminate $\Delta P_{GS}$ and $\Delta P_{GA}$. If the power of the line loss part is ignored, then:
$$\lambda = \pm\sqrt{-d(\Delta P_C + \Delta P_{L2})/d\delta_2} = \pm\sqrt{-df(\delta)/d\delta}. \tag{C5}$$

Thus, if $df(\delta)/d\delta<0$, the eigenvalue has a positive real part, and the equilibrium point is unstable. When $df(\delta)/d\delta=0$, the system has two zero eigenvalues, which are critically stable. When $df(\delta)/d\delta>0$, the system has two imaginary eigenvalues. Since the damping is neglected in the above analysis process, the system is statically stable under positive damping at this time. Therefore, the necessary condition for the stability of the system equilibrium point is $df(\delta)/d\delta>0$.

APPENDIX D

$$f(\delta) = \Delta P_{L2}(\delta) + \Delta P_C(\delta), \tag{D1}$$
where
$$\Delta P_{L2} = P_{L2c} + P_{L2\max}\cos(\delta+\gamma_{L2}),$$
$$\Delta P_C = P_{Cc} + P_{SA\max}\cos(\delta+\gamma_C),$$
$$\Delta P_{L2}(0) = 0, \; \Delta P_C(0) = 0, \; \Delta P_C(-2\gamma_C) = 0.$$
Thus,
$$f(0) = 0, \tag{D2}$$

When $\gamma_C \in (-\pi+|\gamma_{L2}|, 0)$ or $(|\gamma_{L2}|, \pi)$, and $P_{L2\max}>0$, the following formula holds:
$$P_{L2\max}(\cos(-2\gamma_C + \gamma_{L2}) - \cos\gamma_{L2}) < 0. \tag{D3}$$

Since $P_{L2c} = -P_{L2\max}\cos\gamma_{L2}$, therefore,
$$\begin{aligned}f(-2\gamma_C) &= \Delta P_{L2}(-2\gamma_C) \\ &= P_{L2c} + P_{L2\max}\cos(-2\gamma_C + \gamma_{L2}) \\ &= P_{L2\max}(\cos(-2\gamma_C + \gamma_{L2}) - \cos\gamma_{L2}) < 0\end{aligned} \tag{D4}$$

(i) If $\gamma_C \in (-\pi+|\gamma_{L2}|, 0)$ and
$$f'(0) > 0, \tag{D5}$$
there exists at least one zero point in the interval $(0, -2\gamma_C)$ according to the zero point theorem because of Eqs. (D2), (D4), and (D5).